# The Neurobiological Craving Signature (NCS) predicts social craving and responds to social isolation

Ana Defendini[1], Livia Tomova[2,3,4], & Leonie Koban[1,5]*

[1] Lyon Neuroscience Research Center (CRNL), CNRS, Inserm, Université Claude Bernard Lyon 1, Bron, France
[2] Cardiff University Brain Research Imaging Centre (CUBRIC), Cardiff University, UK
[3] Cardiff University Centre for Human Developmental Science (CUCHDS), Cardiff University, UK
[4] Department of Psychology, University of Cambridge, UK
[5] Le Vinatier Psychiatrie Universitaire Lyon Métropole, France

*Please address correspondence to:
Dr. Leonie Koban, CRNL, Institut des Épilepsies IDEE, 59 Boulevard Pinel, 69500 Bron, France;
**Email**: leonie.koban@cnrs.fr

**This PDF file includes:**

    Main Text
    Figures 1 to 2

**Abstract**

Humans are inherently social and seek connection with others for survival. Recent studies suggest that acute social isolation leads to craving for social interactions, but the brain mechanisms of *social craving* and their relationship to brain networks underlying drug and food craving remain incompletely understood. Here we harnessed an existing dataset and tested whether the *Neurobiological Craving Signature* (NCS)—a recently developed fMRI-based brain-signature of drug and food craving—also predicts social craving. During fMRI, participants rated their craving for images of food, social interactions, and flowers in three different sessions: after 10h of fasting from food, 10h of social isolation, or neither (baseline; order of sessions counterbalanced). The NCS significantly predicted self-reported craving for food and social cues but not flower cues. Further, NCS responses to food were higher after fasting compared to baseline, and higher for social cues after social isolation compared to baseline, demonstrating its responsiveness to both food and social deprivation. These findings resonate with recent work showing shared brainstem circuits for hunger and social isolation, and indicate shared whole-brain circuits for social, food, and drug craving. They open new avenues for testing the NCS across different primary rewards, for assessing the consequences of their deprivation, and for examining how social deprivation—such as loneliness and isolation—interacts with overeating and drug use.

**Introduction**

Humans are social animals and depend on others for survival (1–3). Lack of social connection can have substantial consequences for health and is associated with increased mortality (4, 5). Correspondingly, feelings of loneliness arise when social needs are unmet and constitute a risk factor for several mental health disorders including anxiety, depression (6–9), addiction (10, 11), and eating disorders (12, 13). Understanding the neurobiological mechanisms underlying the human motivation for social connection and the consequences of social isolation—an objective lack of social contact and interaction—would pave the way for more effective prevention and treatment strategies for mental and bodily health conditions associated with isolation and loneliness.

Previous neuroimaging studies have shown that social stimuli such as faces are rewarding and increase activity in areas that also respond to other primary rewards (e.g., food) and secondary rewards (e.g., monetary) (14–17). Similarly, anticipation of social rewards, such as attractive faces, erotic stimuli, expressions of positive affect, and abstract signals of social approval, has been shown to engage brain areas that overlap with those observed during anticipation of drug and food rewards (16, 18–20). However, less is known about how the brain processes more complex social stimuli, such as observed interactions between others (21, 22), how brain responses to social stimuli relate to self-reported motivation for social contact, and how brain and self-reported motivational responses to social stimuli are shaped by states of social isolation or loneliness (23). Recent findings show that, in healthy, socially connected young adults, a short 10h period of social isolation produces similar response patterns as 10h of food fasting in the midbrain, encompassing the substantia nigra (SN) and the ventral tegmental area (VTA)(24). These results have led to the notion of '*social craving*' which can be defined as the desire for meaningful interactions and relationships with others. Yet, it remains unclear whether *social craving* is predicted by the same whole brain networks as those that predict craving in the context of addiction or eating disorders, where it is defined as the strong urge or desire to use a drug or to eat food. Here, we harness the *Neurobiological Craving Signature* (NCS)—a recently developed brain marker that predicts the intensity of drug and food craving and separates drug users from matched controls (25)—to test whether craving for social interactions is predicted by the same brain patterns that predict craving for drugs and highly palatable food.

Social isolation, rejection, and loneliness are known risk factors for mental health problems, dysregulated eating behavior and substance use. More specifically, people who report higher objective social isolation and more subjective feelings of loneliness are more susceptible to suffer from addictions and eating disorders (9, 12). In turn, eating and substance use disorder may often lead to social marginalization and altered sensitivity to social exclusion (26, 27). Shared whole-brain patterns for drug, food, and social craving could potentially explain the bidirectional associations between loneliness and dysregulated behavior for drugs and food. One possibility is that deprivation of one reward is non-specific and increases motivation to seek other, related rewards that engage overlapping neural circuitry (28, 29). Alternatively, overlapping neural patterns may reflect a common motivational state after deprivation that remains selective to the unmet need, consistent with evidence that midbrain responses are specific to a deprived reward cue (24). Here we test these possibilities by assessing whether and how the NCS responds to social and food deprivation.

While *social craving* is a novel term, drug and food craving have been extensively studied in the context of substance use and eating disorders (30, 31), for which craving is a core symptom. Drug and food craving are known as strong predictors of relapse, substance use, and binge eating episodes, respectively (32, 33). Neuroimaging studies have used cue-reactivity paradigms to show that different types of legal and illicit drugs, appetizing foods, and cues for behavioral addictions (such as gambling) evoke responses in dopaminergic brain regions implicated in valuation, reward, and affective processing (34–37). Recently, this type of paradigm has been combined with a machine-learning approach to develop a *brain signature* of drug and food craving—the Neurobiological Craving Signature (NCS)—that predicts craving based on distributed brain activation patterns. Brain signatures are brain pattern-based predictive models that decode mental processes based on multivariate, distributed brain activity (38–40). Brain signatures aim to provide sensitive and specific brain-based measures of specific mental processes at the population level and have been developed previously for processes such as pain (41), monetary rewards (42), and imagination (43). The Neurobiological Craving Signature (NCS) was developed to predict the intensity of drug and food craving and discriminates drug users versus non-users with >80% out-of-sample accuracy (25). Thus, the NCS captures a distributed pattern of activation across the whole brain that predicts self-reported craving across individuals. It has positive weights in areas related to reward and motivation, but also significant weights in other areas such as lateral prefrontal and parietal areas. When the NCS is applied to novel fMRI data, higher NCS responses indicate stronger expression of this neurobiological craving pattern and thus higher activation of craving-related brain regions, whereas lower responses indicate weaker expression. If indeed "*social craving*"—the motivational drive for social interactions—shares common neurophysiological mechanisms with drug and food craving, then the NCS—developed to predict drug and food craving in a completely independent sample of drug users and matched controls—should also be able to predict the degree of self-reported social craving and respond to an experimentally induced deprivation of social needs.

In this study, we tested the NCS on an existing fMRI dataset (44) to address this question. All participants were scanned three times: once after 10h without any access to food but with access to social interactions (fasting condition), once after 10h of social isolation with no access to in-person, phone, or online social interactions but with access to food (social isolation condition), and once after a baseline session where participants did not undergo any fasting or social isolation (baseline condition). The order of session was counterbalanced across participants. During fMRI, participants performed a cue-induced craving task (Figure 1a) to measure brain responses and block-wise craving ratings for food and social cues, and for positive but non-craving related stimuli (flowers). By applying the NCS to this dataset, we tested several novel research questions, namely 1) whether the NCS predicted social craving ratings across blocks, which would suggest shared whole-brain patterns for drug, food, and social craving, and 2) whether NCS responses were higher for food cues and social cues after respectively fasting from food and abstaining from social interactions for 10h, which would indicate the sensitivity of the NCS to short-term deprivation and support its validity as a common brain target to measure the effects of experimental interventions on both food and social craving.

**Results**

***The NCS predicts self-reported craving of food and social interactions***

We first tested the hypothesis that NCS responses predict self-reported craving for food and for social cues. The cue-induced craving task consisted of blocks of three images of the same category (food, social interactions, or flowers). After each block, participants rated on a 0-10 scale how much they craved food, social interactions, or how much they liked flowers, respectively. We used a general linear model (GLM) to model brain responses separately for each block, and then applied the NCS to the resulting beta images for each participant, by computing the dot product between the NCS weight map and the beta image (c.f.,(25)) and resulting in one scalar NCS response per block. For each cue category (food, social, and flower), we then computed multi-level GLMs where the NCS response was used as a predictor of self-reported craving ratings across all sessions (baseline, isolation and fasting). In line with the previous training of the NCS on an independent data set employing drug and food cues (25), NCS responses significantly predicted self-reported craving for food cues (Fig. 1c, $t(29) = 2.35$, $p = 0.026$, Cohen's $d = 0.42$), validating the NCS as a neuromarker of food craving, even in new samples and different tasks. Further, and in line with our hypothesis, NCS responses significantly predicted self-reported craving for social cues ($t(29) = 2.16$, $p = 0.039$, $d = 0.37$, see Fig. 1c), demonstrating that social craving is predicted by the same whole-brain activity pattern that was trained to predict drug and food craving. In contrast, the NCS did not predict ratings of flower cues ($t(29) = -0.26$, $p = 0.798$), showing that the NCS does not predict liking or pleasantness in general, but specifically predicts craving (Fig 1c).

### *NCS responses to food cues are higher after fasting and higher for social cues after social isolation*

Second, we tested the effects of experimental food deprivation and social isolation on NCS responses (Fig. 2) We performed one-way ANOVAs to compare the effect of session (fasting, baseline, and isolation) on craving for the three types of cues (food, social, flowers). We found significant effects of session on NCS responses to food (Fig. 2a) ($F$(1.92, 55.72) = 3.51, $p$ = 0.038, generalized eta squared ($η2$ = 0.031) and social cues (Fig. 2b) ($F$(1.95, 53.62) = 3.89, $p$ = 0.029, $η2$= 0.033), but not for NCS response to flower cues (Fig. 2c) ($F$(1.84, 53.45) = 2.18, $p$ = 0.126, $η2$= 0.012)). Planned comparisons (pairwise t-test) showed that, in line with our hypothesis, NCS response to food cues were significantly higher after the fasting than after the baseline session (CI: [0.12, 1.28], $t$ (29) = 2.46, $p$ = 0.020, $d$ = 0.45). Paralleling these effects, NCS responses to social cues were significantly higher after social isolation compared to the baseline session (CI: [0.17, 1.57], $t$(29) = 2.53, $p$ = 0.017, $d$ = 0.46), showing that NCS responses selectively increase for the deprived stimulus type.

**Discussion**

Social contact is a basic human need (1–3), but less is known about how the brain regulates motivation for social interactions, especially in humans (14, 45). Recent work shows that social isolation elicits midbrain responses to social cues that resemble responses to food cues following fasting (24). Building on these findings, we tested the hypothesis that craving for social contact may engage similar whole-brain mechanisms as drug and food craving. For this purpose, we harnessed the Neurobiological Craving Signature (NCS) (25), which has been developed previously using completely independent datasets to predict drug and food craving based on whole-brain activity patterns. Our results show that the NCS predicts both food and social craving, but not liking of flowers, demonstrating shared brain patterns underlying drug, food, and social craving. We also provide first evidence that the NCS is modulated in stimulus-specific ways by short-term deprivation: NCS responses to food cues were increased after 10h of fasting, and NCS responses to social cues were increased after 10h of social isolation. These findings do not only extend recent findings from the same dataset on midbrain responses to social and food deprivation (24), but provide several novel insights regarding the functional neurobiology of craving on a whole-brain level, discussed in more detail below. Together, they confirm the concept of 'social craving' and provide evidence that the NCS can be fruitfully applied to study social craving. Given that social isolation, marginalization, and subjective feelings of loneliness increase the risk of substance use, overeating, and other forms of addiction (30, 46–49), the NCS could be harnessed in future studies to investigate the potential interactions between drug, food, and social craving, and their role in substance use and eating disorders (50).

### *Shared whole brain patterns for social, drug, and food craving*

First, our results confirm that the NCS predicts the intensity of self-reported food craving, in line with the development of the NCS on drug and food stimuli (25). This finding demonstrates the validity of the NCS's in predicting food craving in a completely new dataset, from a different group of investigators, and which used a different task design (e.g., block design instead of event-related design) and different form of rating. It thereby highlights the generalizability of the NCS as a neuromarker of food craving across different experimental settings and the value of brain signatures as formal and falsifiable brain models for the robustness and reproducibility of neuroimaging research (38).

Second, we found that the NCS generalizes to social craving. It predicts the intensity of self-reported craving for social interactions, supporting the hypothesis that drug, food, and social craving share similar neurobiological processes on the whole-brain level. This finding resonates with previous work that has shown similar responses of reward circuits for social and other rewards in both humans and other animals (51–54) and confirm previous results from the same datasets of shared pattern for hunger and social isolation in the midbrain (24). The present findings, testing an independently established brain signature, provides the most conclusive evidence to date for shared mesoscale whole-brain patterns that predict the intensity of social, drug, and food craving. In contrast to previous work, they do not only identify shared patterns on the level of individual subjects, but on the population level. This opens the possibility of further testing the NCS as a neuromarker for social craving in future studies, for example by investigating how social craving may be altered in neuropsychiatric and neurodevelopmental conditions, such as depression, social anxiety, or autism spectrum disorder, where processing of social rewards processing and other social cognitive processes are often affected (25, 55–57).

The NCS predicted craving for food and social interactions but did not predict liking of flowers in this dataset. This demonstrates the specificity and discriminative validity of the NCS in predicting craved and motivationally relevant stimuli, as opposed to liking or pleasantness of stimuli. This is in line with previous suggestions that 'liking' and 'wanting' are similar but dissociable psychological processes (58). For instance, whereas 'craving' has been associated with increased mesolimbic dopamine activity in response to reward-predicting cues, 'liking' has been linked to the hedonic aspect of rewards that involve opioid-sensitive hedonic hotspots within the nucleus accumbens and ventral pallidum (59, 60). Future studies could compare the NCS and pleasure-related brain signatures (60) to distinguish between 'liking' and 'wanting' processes, and continue to assess the generalizability of the NCS to other forms of craving. For example, future studies could test whether the NCS differs between primary (e.g., food, social) and secondary (e.g., monetary) rewards (52, 61), including in the context of behavioral addictions such as gambling disorder.

### *The NCS is sensitive to deprived reward*

Third, our results demonstrate that food and social deprivation modulate NCS responses specifically for the deprived stimulus. NCS responses to food (but not to social or flower) cues were increased following 10h of fasting compared to baseline. NCS responses to social (but not food or flower) cues were increased following 10h of social isolation compared to baseline. This further highlights the validity of the NCS as a neuromarker of both food and social craving, and its responsiveness to experimental manipulations designed to alter craving and motivational states, making it an interesting target for intervention studies.

The effects of session (Fasting vs. Isolation vs. Baseline) on the NCS yielded some subtle but interesting differences as compared with the regions-of-interest (ROI) approach used by Tomova et al. 2020 (24) in the SN/VTA midbrain regions that are important for reward expectation and incentive motivation (62, 63). In the present analysis, NCS responses were *increased* following deprivation compared to baseline (and non-significantly but numerically compared to the other type of deprivation). In contrast, Tomova et al. 2020 (24) found that activity in the SN/VTA (and/or a functionally defined ROI) was *decreased* for the nondeprived need (rather than increased for the deprived need), suggesting a narrowing of reward-sensitive midbrain circuits following deprivation. This might reflect a functional difference between local, midbrain responses versus whole-brain responses to craving. Whereas midbrain structures such as the SN/VTA may have a role in 'gating' relevant stimuli, by suppressing responses to irrelevant (in this case, non-deprived) stimuli, NCS responses, which integrate activity patterns across the whole brain, may be closer to the more complex and subjective experience of craving, and thus reflect a combination of gating and motivational, but also higher-order cognitive processes (63–66).

In addition, several subcortical and cortical ROIs examined by Tomova et al., 2020 (24) showed differences between deprivation-related brain responses for food and social cues. For instance, the nucleus accumbens (NAc) and putamen showed higher responses to food versus flower cues after fasting, while the caudate showed higher responses to social versus flower cues after isolation. Moreover, a whole-brain conjunction analysis between food cue related responses after fasting and social cue related responses after isolation did not reveal any significant overlap. On first sight, this might seem at odds with our finding that NCS response increases for both food and social cues following cue-specific deprivation compared to baseline conditions. However, shared NCS responses to social, drug, and food craving do not necessarily preclude domain-specific responses on more local levels of analyses. Instead, they are in line with the idea of a domain-general network of craving that integrates different subprocesses, which might somewhat vary in their relative importance across domains (and between individuals). Further, by integrating information across large-scale, multivariate patterns, whole brain signatures yield greater statistical power than mass-univariate brain analyses, which need to correct for multiple comparisons across thousands of voxels (38, 39, 67, 68). We also note that, while the NCS responded to both food and social deprivation, it did so in a cue-specific way, with higher NCS responses to food (but not social) cues following fasting, and higher NCS responses to social (but not food) cues after isolation. However, we note that during social isolation participants had ad libitum access to food and that this study was not designed to detect potential interactions between social and food craving, which could be explored in future research. Future work could also use the NCS in combination with ROI analyses to further characterize the differences between domain specific and domain general craving, and the effects of deprivation on the neurobiological mechanisms.

The data of 30 participants across the 3 different scanning sessions (corresponding to a total of 90 fMRI sessions) was available for the present analysis. Future studies could test social craving in larger sample size to explore potential sex and gender differences, as well as the role of individual differences in brain responses underlying social and other forms of craving. Future studies could also test NCS and brain responses to other modalities of social stimuli (e.g., voices, laughter, social touch) after social isolation and investigate how social craving might be differentially expressed in clinical conditions (e.g., substance use disorders (69)) and across different developmental phases.

In sum, the present findings suggest an important convergence of the brain patterns underlying drug, food, and social craving in humans, and highlight the usefulness of a brain signature approach for testing specific functional hypotheses. Future studies could harness the NCS to advance our knowledge on social craving and its interactions with other types of craving. For instance, they could apply the NCS to fMRI studies measuring social craving after mandatory isolation phases (i.e., as in the case during the COVID-19 pandemic) and use the NCS response to observe how craving for social interactions relates to feelings of loneliness, especially among young adults and elderly individuals (7, 70, 71). Additionally, the NCS could serve to test whether psychosocial interventions can help to regulate social craving and mitigate the effects of isolation (9, 72). Given its sensitivity to drug, food, and social craving, the NCS is also a promising target to advance our understanding of the neurobiological mechanisms that underlie the interactions between social isolation or marginalization, overeating, and drug use in at-risk and clinical populations (73–76).

**Materials and Methods**

*Dataset*

We used a previously published dataset (24), available on the open-source website OpenNeuro (identifier ds003242 (44)), with additional behavioral data provided by the authors (LT), to perform separate and novel analyses using the Neurobiological Craving Signature (NCS) (25). The dataset on OpenNeuro contained fMRI data from 32 healthy adult volunteers. Two participants were excluded from this dataset due to incomplete or missing functional runs. Therefore, our final analysis used data from 30 healthy adults (mean age = 27; 20 female). All participants provided informed consent and were scanned while they performed a cue-induced craving task in three separate sessions on different days. Anatomical and functional scans were acquired using a Siemens Prisma 3T scanner. T1*weighted structural images were collected with 1mm isotropic voxels (256 FOV) in 176 interleaved sagittal slices. Functional data was collected in six runs of 147 volumes with 58 T2* weighted echo planar slices with the following parameters: TR=2s, TE=30ms, FOV=210mm, 3x3x3mm$^3$ voxel size).

*Experimental procedures*

Details regarding the design and the data acquisition of the original study were previously reported (24). In brief, participants completed three experimental sessions on different days, corresponding to three different experimental condition (order counterbalanced across participants): Baseline, Fasting, and Social Isolation. In the Baseline condition, participants assisted the scanning session without any previous intervention. In the Fasting condition, participants had to abstain from eating and drinking anything else than water for 10h before the MRI session. In the Isolation condition, participants were socially isolated for 10h before the MRI session (no access to in-person interactions, phone calls, social media). During scanning, participants completed a cue-induced craving paradigm during which participants were shown images of 1) highly palatable foods (e.g., chocolate, pizza), 2) positive social interactions (e.g., groups of individuals engaging with each other, talking, laughing) and 3) a control condition (attractive flowers). These items were individualized and selected based on each participant's favorite foods and social activities. In each session, participants were presented with 12 blocks (36 images) per cue condition. Different images were presented for each session, resulting in a total of 36 blocks and 108 images per cue condition across all three sessions. Each trial displayed an image with a short 3–5-word description of the image, displayed for 5s, and was followed by a 1s fixation cross. Blocks (consisting of three trials of the same cue category) were followed by a 2-6s jittered period and a self-reported measure of how much participants were craving the food or social interaction, or how much they liked the flower image. Following the ratings, participants saw another jittered fixation cross period of 2-6s before the onset of the next block.

*fMRI preprocessing and analysis*

SPM12 was used to preprocess the MRI data. Functional images were realigned to correct for motion artifacts, followed by slice-timing correction. Anatomical scans were co-registered to the mean functional image and segmentation of the anatomical data was performed using tissue probability maps. The functional and anatomical images were normalized to MNI space with a 2mm isotropic voxel size. A Gaussian kernel of a full width at half maximum (FWHM) of 4mm was used for smoothing and a high-pass filter with a cutoff of 128 seconds was applied to remove low-frequency noise from the data. For each session (baseline, fasting and isolation), we computed a first-level general linear model (GLM) for every participant to estimate brain (BOLD) responses to the cues. The design matrix included separate regressors for each stimulus block in the three conditions (food, social and flower), as well as six movement regressors of no interest to control for motion artefacts. This allowed us to compute beta images for each of the 12 blocks per condition and session for each participant.

*Statistical analyses*

NCS responses were calculated as the dot-product between the L2-normed individual beta-images for each cue block and the NCS weight map, resulting in 108 NCS scalar responses per participant (12 x 3 cue types x 3 conditions). The relationship between NCS responses and self-reported craving was statistically analyzed using multilevel GLM (code available on https://github.com/canlab/CanlabCore) for each cue condition (food, social and flower) separately. Multilevel GLM enables the use of participants as a random effect by first implementing a random-slope model across participants fitting their regressions individually as $2^{nd}$-level units and then uses a precision-weighted least squares approach to model the group effects.

Further, to compare differences in NCS responses for each condition (food, flower and social interaction cues) across the three experimental sessions (baseline, fasting and isolation), we averaged the NCS response across the 12 blocks for each condition and session per participant. We then used one-way ANOVAs to test whether the session type had any significant effect on NCS responses within each cue type and performed two sided paired-wise t-tests. All analyses used a two-sided threshold of $p < .05$ to determine significant differences. Cohen's *d* was calculated to determine the effect sizes of pairwise t-tests.

**Acknowledgments**

This work was funded by an ERC Starting Grant to LK (101041087). Views and opinions expressed are however those of the authors only and do not necessarily reflect those of the European Union or the European Research Council. Neither the European Union nor the granting authority can be held responsible for them. The funders had no role in study design, data analysis, manuscript preparation, or publication decisions.

**Data and code availability**

Data is publicly available on OpenNeuro (identifier ds003242). The NCS weight map and code to apply it is available on GitHub (https://github.com/ldmk/NCS).

**References**

1. P. Chen, W. Hong, Neural Circuit Mechanisms of Social Behavior. *Neuron* **98**, 16–30 (2018).
2. R. Adolphs, The neurobiology of social cognition. *Current Opinion in Neurobiology* **11**, 231–239 (2001).
3. R. F. Baumeister, M. R. Leary, The need to belong: Desire for interpersonal attachments as a fundamental human motivation. *Psychological Bulletin* **117**, 497–529 (1995).
4. J. S. House, K. R. Landis, D. Umberson, Social Relationships and Health. *Science* **241**, 540–545 (1988).
5. J. Holt-Lunstad, Why Social Relationships Are Important for Physical Health: A Systems Approach to Understanding and Modifying Risk and Protection. *Annu. Rev. Psychol.* **69**, 437–458 (2018).
6. J. T. Cacioppo, M. E. Hughes, L. J. Waite, L. C. Hawkley, R. A. Thisted, Loneliness as a specific risk factor for depressive symptoms: Cross-sectional and longitudinal analyses. *Psychology and Aging* **21**, 140–151 (2006).
7. J. Holt-Lunstad, Loneliness and Social Isolation as Risk Factors: The Power of Social Connection in Prevention. *American Journal of Lifestyle Medicine* **15**, 567–573 (2021).
8. N. Leigh-Hunt, *et al.*, An overview of systematic reviews on the public health consequences of social isolation and loneliness. *Public Health* **152**, 157–171 (2017).
9. J. Holt-Lunstad, T. B. Smith, M. Baker, T. Harris, D. Stephenson, Loneliness and Social Isolation as Risk Factors for Mortality: A Meta-Analytic Review. *Perspect Psychol Sci* **10**, 227–237 (2015).
10. M. Heilig, D. H. Epstein, M. A. Nader, Y. Shaham, Time to connect: bringing social context into addiction neuroscience. *Nat Rev Neurosci* **17**, 592–599 (2016).
11. B. E. Walsh, R. C. Schlauch, Differential impact of emotional and social loneliness on daily alcohol consumption in individuals with alcohol use disorder. *Drug and Alcohol Dependence* **264**, 112433 (2024).
12. M. P. Levine, Loneliness and Eating Disorders. *The Journal of Psychology* **146**, 243–257 (2012).
13. L. Cortés-García, R. Rodríguez-Cano, T. Von Soest, Prospective associations between loneliness and disordered eating from early adolescence to adulthood. *Intl J Eating Disorders* **55**, 1678–1689 (2022).
14. L. Tomova, K. Tye, R. Saxe, The neuroscience of unmet social needs. *Social Neuroscience* **16**, 221–231 (2021).
15. U. J. Pfeiffer, *et al.*, Why we interact: On the functional role of the striatum in the subjective experience of social interaction. *NeuroImage* **101**, 124–137 (2014).
16. K. Izuma, D. N. Saito, N. Sadato, Processing of Social and Monetary Rewards in the Human Striatum. *Neuron* **58**, 284–294 (2008).
17. J. P. Bhanji, M. R. Delgado, The social brain and reward: social information processing in the human striatum. *WIRES Cognitive Science* **5**, 61–73 (2014).
18. I. Aharon, *et al.*, Beautiful Faces Have Variable Reward Value. *Neuron* **32**, 537–551 (2001).
19. C. C. Ruff, E. Fehr, The neurobiology of rewards and values in social decision making. *Nat Rev Neurosci* **15**, 549–562 (2014).
20. N. D. Volkow, G. J. Wang, J. S. Fowler, D. Tomasi, R. Baler, "Food and Drug Reward: Overlapping Circuits in Human Obesity and Addiction" in *Brain Imaging in Behavioral Neuroscience*, Current Topics in Behavioral Neurosciences., C. S. Carter, J. W. Dalley, Eds. (Springer Berlin Heidelberg, 2011), pp. 1–24.
21. M. Iacoboni, *et al.*, Grasping the Intentions of Others with One's Own Mirror Neuron System. *PLoS Biol* **3**, e79 (2005).
22. J. Sliwa, W. A. Freiwald, A dedicated network for social interaction processing in the primate brain. *Science* **356**, 745–749 (2017).
23. S. Cacioppo, J. P. Capitanio, J. T. Cacioppo, Toward a neurology of loneliness. *Psychological Bulletin* **140**, 1464–1504 (2014).
24. L. Tomova, *et al.*, Acute social isolation evokes midbrain craving responses similar to hunger. *Nat Neurosci* **23**, 1597–1605 (2020).

25. L. Koban, T. D. Wager, H. Kober, A neuromarker for drug and food craving distinguishes drug users from non-users. *Nat Neurosci* **26**, 316–325 (2023).
26. P. Bach, *et al.*, Effects of social exclusion and physical pain in chronic opioid maintenance treatment: fMRI correlates. *European Neuropsychopharmacology* **29**, 291–305 (2019).
27. P. Maurage, *et al.*, Disrupted Regulation of Social Exclusion in Alcohol-Dependence: An fMRI Study. *Neuropsychopharmacol* **37**, 2067–2075 (2012).
28. K. C. Berridge, T. E. Robinson, Liking, wanting, and the incentive-sensitization theory of addiction. *American Psychologist* **71**, 670–679 (2016).
29. G. Sescousse, Y. Li, J.-C. Dreher, A common currency for the computation of motivational values in the human striatum. *Social Cognitive and Affective Neuroscience* **10**, 467–473 (2015).
30. M. A. Sayette, The Role of Craving in Substance Use Disorders: Theoretical and Methodological Issues. *Annu. Rev. Clin. Psychol.* **12**, 407–433 (2016).
31. M. L. Pelchat, Of human bondage. *Physiology & Behavior* **76**, 347–352 (2002).
32. N. Vafaie, H. Kober, Association of Drug Cues and Craving With Drug Use and Relapse: A Systematic Review and Meta-analysis. *JAMA Psychiatry* **79**, 641 (2022).
33. R. G. Boswell, H. Kober, Food cue reactivity and craving predict eating and weight gain: a meta-analytic review. *Obesity Reviews* **17**, 159–177 (2016).
34. W. Sun, H. Kober, Regulating food craving: From mechanisms to interventions. *Physiology & Behavior* **222**, 112878 (2020).
35. D. W. Tang, L. K. Fellows, D. M. Small, A. Dagher, Food and drug cues activate similar brain regions: A meta-analysis of functional MRI studies. *Physiology & Behavior* **106**, 317–324 (2012).
36. A. Schienle, A. Schäfer, A. Hermann, D. Vaitl, Binge-Eating Disorder: Reward Sensitivity and Brain Activation to Images of Food. *Biological Psychiatry* **65**, 654–661 (2009).
37. M. L. Pelchat, A. Johnson, R. Chan, J. Valdez, J. D. Ragland, Images of desire: food-craving activation during fMRI. *NeuroImage* **23**, 1486–1493 (2004).
38. P. A. Kragel, L. Koban, L. F. Barrett, T. D. Wager, Representation, Pattern Information, and Brain Signatures: From Neurons to Neuroimaging. *Neuron* **99**, 257–273 (2018).
39. C.-W. Woo, L. J. Chang, M. A. Lindquist, T. D. Wager, Building better biomarkers: brain models in translational neuroimaging. *Nat Neurosci* **20**, 365–377 (2017).
40. J. D. E. Gabrieli, S. S. Ghosh, S. Whitfield-Gabrieli, Prediction as a Humanitarian and Pragmatic Contribution from Human Cognitive Neuroscience. *Neuron* **85**, 11–26 (2015).
41. T. D. Wager, *et al.*, An fMRI-Based Neurologic Signature of Physical Pain. *N Engl J Med* **368**, 1388–1397 (2013).
42. S. P. H. Speer, *et al.*, A multivariate brain signature for reward. *NeuroImage* **271**, 119990 (2023).
43. S. Lee, *et al.*, A neural signature of the vividness of prospective thought is modulated by temporal proximity during intertemporal decision making. *Proc. Natl. Acad. Sci. U.S.A.* **119**, e2214072119 (2022).
44. L. Tomova, *et al.*, MRI data of 40 adult participants in response to a cue induced craving task following food fasting, social isolation and baseline (within-subject design). Openneuro. https://doi.org/10.18112/OPENNEURO.DS003242.V1.0.0. Deposited 2020.
45. J. Cressy, C. Jia, J. Salk, K. M. Tye, Neural Mechanisms of Social Homeostasis: Dynamic Range Plasticity. *J. Neurosci.* **46**, e0224252025 (2026).
46. M. Heilig, *et al.*, Addiction as a brain disease revised: why it still matters, and the need for consilience. *Neuropsychopharmacol.* **46**, 1715–1723 (2021).
47. N. C. Christie, The role of social isolation in opioid addiction. *Social Cognitive and Affective Neuroscience* **16**, 645–656 (2021).
48. A. Y. Farmer, Y. Wang, N. A. Peterson, S. Borys, D. K. Hallcom, Social Isolation Profiles and Older Adult Substance Use: A Latent Profile Analysis. *The Journals of Gerontology: Series B* **77**, 919–929 (2022).
49. K. Hanna, J. Cross, A. Nicholls, D. Gallegos, The association between loneliness or social isolation and food and eating behaviours: A scoping review. *Appetite* **191**, 107051 (2023).

50. W. Zheng, *et al.*, Effect of Natural Rewards on Substance Use Disorder: An Incentive Sensitization Perspective. *Biological Psychiatry* **99**, 436–445 (2026).
51. F. Rocha-Almeida, A. R. Conde-Moro, A. Fernández-Ruiz, J. M. Delgado-García, A. Gruart, Cortical and subcortical activities during food rewards versus social interaction in rats. *Sci Rep* **15**, 4389 (2025).
52. G. Sescousse, X. Caldú, B. Segura, J.-C. Dreher, Processing of primary and secondary rewards: A quantitative meta-analysis and review of human functional neuroimaging studies. *Neuroscience & Biobehavioral Reviews* **37**, 681–696 (2013).
53. K. C. Berridge, M. L. Kringelbach, Pleasure Systems in the Brain. *Neuron* **86**, 646–664 (2015).
54. G. A. Matthews, K. M. Tye, Neural mechanisms of social homeostasis. *Annals of the New York Academy of Sciences* **1457**, 5–25 (2019).
55. Z. I. Santini, A. Koyanagi, S. Tyrovolas, C. Mason, J. M. Haro, The association between social relationships and depression: A systematic review. *Journal of Affective Disorders* **175**, 53–65 (2015).
56. T. Takumi, K. Tamada, F. Hatanaka, N. Nakai, P. F. Bolton, Behavioral neuroscience of autism. *Neuroscience & Biobehavioral Reviews* **110**, 60–76 (2020).
57. S. Pearcey, *et al.*, Research Review: The relationship between social anxiety and social cognition in children and adolescents: a systematic review and meta-analysis. *Child Psychology Psychiatry* **62**, 805–821 (2021).
58. K. C. Berridge, Pleasures of the brain. *Brain and Cognition* **52**, 106–128 (2003).
59. K. S. Smith, K. C. Berridge, Opioid Limbic Circuit for Reward: Interaction between Hedonic Hotspots of Nucleus Accumbens and Ventral Pallidum. *J. Neurosci.* **27**, 1594–1605 (2007).
60. P. A. Kragel, M. T. Treadway, R. Admon, D. A. Pizzagalli, E. C. Hahn, A mesocorticolimbic signature of pleasure in the human brain. *Nat Hum Behav* **7**, 1332–1343 (2023).
61. B. Knutson, G. W. Fong, C. M. Adams, J. L. Varner, D. Hommer, Dissociation of reward anticipation and outcome with event-related fMRI: *Neuroreport* **12**, 3683–3687 (2001).
62. J.-H. Baik, Dopamine signaling in food addiction: role of dopamine D2 receptors. *BMB Reports* **46**, 519–526 (2013).
63. N. D. Volkow, M. Michaelides, R. Baler, The Neuroscience of Drug Reward and Addiction. *Physiological Reviews* **99**, 2115–2140 (2019).
64. E. P. Merikle, The Subjective Experience of Craving: An Exploratory Analysis. *Substance Use & Misuse* **34**, 1101–1115 (1999).
65. J. C. Horvitz, Dopamine gating of glutamatergic sensorimotor and incentive motivational input signals to the striatum. *Behavioural Brain Research* **137**, 65–74 (2002).
66. B. Lloyd, *et al.*, Subcortical nuclei of the human ascending arousal system encode anticipated reward but do not predict subsequent memory. *Cerebral Cortex* **35**, bhaf101 (2025).
67. M. C. Reddan, M. A. Lindquist, T. D. Wager, Effect Size Estimation in Neuroimaging. *JAMA Psychiatry* **74**, 207 (2017).
68. P. A. Kragel, X. Han, T. E. Kraynak, P. J. Gianaros, T. D. Wager, Functional MRI Can Be Highly Reliable, but It Depends on What You Measure: A Commentary on Elliott et al. (2020). *Psychol Sci* **32**, 622–626 (2021).
69. A. Löfberg, *et al.*, The neurobiological cravings signature (NCS) as a predictive neuromarker of clinical outcomes in alcohol use disorder. *Neuropsychopharmacol.* (2026). https://doi.org/10.1038/s41386-026-02369-3.
70. J. Martino, J. Pegg, E. P. Frates, The Connection Prescription: Using the Power of Social Interactions and the Deep Desire for Connectedness to Empower Health and Wellness. *American Journal of Lifestyle Medicine* **11**, 466–475 (2017).
71. A. Orben, L. Tomova, S.-J. Blakemore, The effects of social deprivation on adolescent development and mental health. *The Lancet Child & Adolescent Health* **4**, 634–640 (2020).
72. F. Mehrabi, F. Béland, Effects of social isolation, loneliness and frailty on health outcomes and their possible mediators and moderators in community-dwelling older adults: A scoping review. *Archives of Gerontology and Geriatrics* **90**, 104119 (2020).

73. S. Song, A. Zilverstand, W. Gui, X. Pan, X. Zhou, Reducing craving and consumption in individuals with drug addiction, obesity or overeating through neuromodulation intervention: a systematic review and meta-analysis of its follow-up effects. *Addiction* **117**, 1242–1255 (2022).
74. M. Xiao, Y. Luo, H. Chen, Social support influences effective neural connections during food cue processing and overeating: A bottom-up pathway. *International Journal of Clinical and Health Psychology* **25**, 100545 (2025).
75. E. L. Gardner, "Addiction and Brain Reward and Antireward Pathways" in *Advances in Psychosomatic Medicine*, M. R. Clark, G. J. Treisman, Eds. (S. Karger AG, 2011), pp. 22–60.
76. R. Sinha, Stress and substance use disorders: risk, relapse, and treatment outcomes. *Journal of Clinical Investigation* **134**, e172883 (2024).

**Figures**

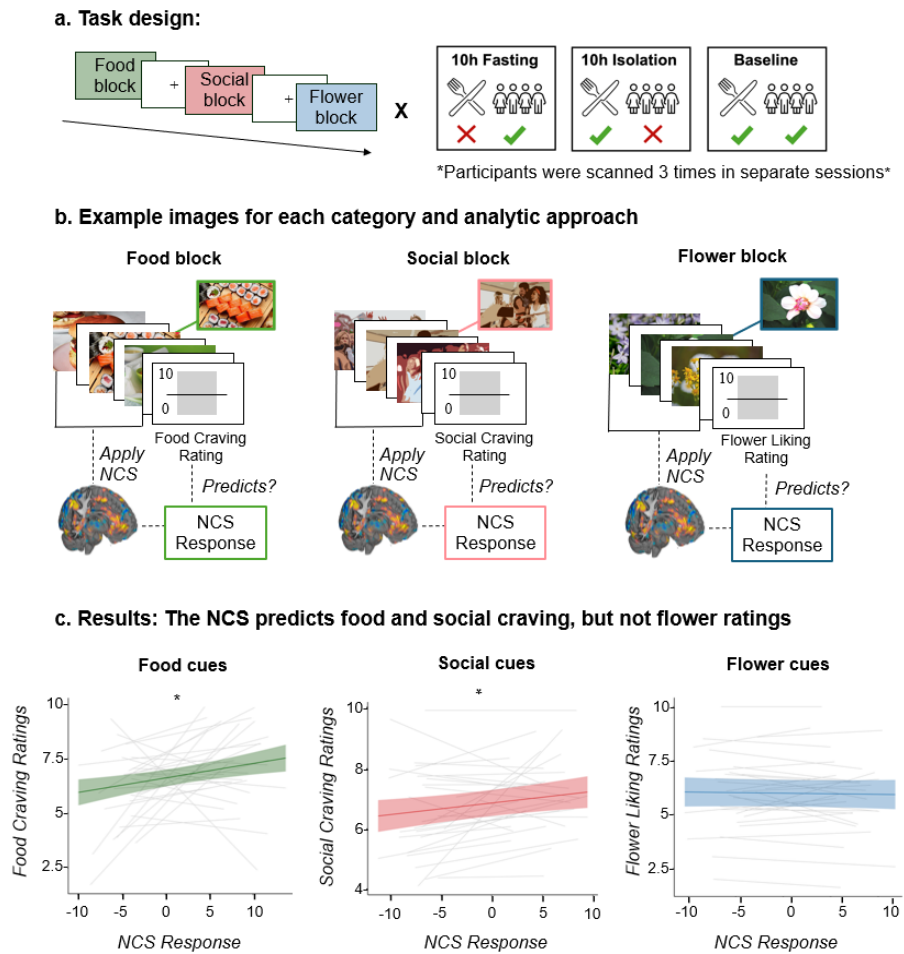

**Figure 1. Task design, analytic approach, and prediction results.** a) Summary of experimental procedures and the cue-induced craving task (adapted from (24)). All participants were scanned three times in separate sessions that were spaced out for 24h or more (session order was counterbalanced across participants). Participants were asked to either fast from food for 10h (Fasting session), abstain from any social interactions for 10h (Isolation session), or not abstain from either (Baseline session). At the end of each session, participants completed a cue-induced craving task during fMRI scanning. They were presented blocked images of food, social, and flower cues and rated how much they wanted those cues (or liked in the case of flower) after each block. b) Example cues and overview of analytic approach. Images were taken from a free photo data with no copyright restrictions. Social cue images were animated to avoid recognizable faces. The NCS response to each block of cues is calculated as the dot product between the NCS weight map and the beta image for this block, producing one scalar value per block and per participant. c) NCS responses significantly predicted food and social craving but not flower ratings. Graphs show the average predicted ratings as a function of NCS response for food, social, and flower cues. Shaded area represents 95% confidence interval for the predicted ratings. The NCS predicted ratings for food cues (in green), $t(29) = 2.35$, $p = 0.026$, Cohen's $d = 0.42$, and for social interactions (in red), $t(29) = 2.16$, $p = 0.039$, $d = 0.37$, but not for flowers (in blue), $t(29) = -0.26$, $p = 0.798$.

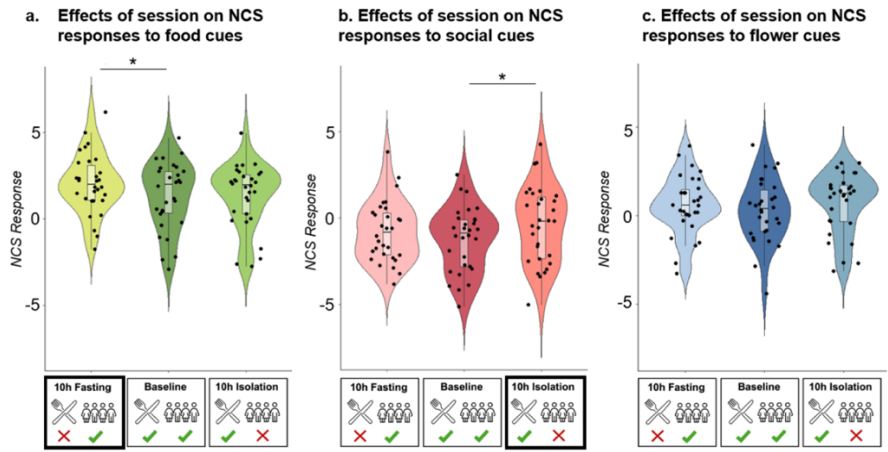

**Figure 2. Effects of fasting and social isolation on NCS responses.** Boxplots show NCS responses for a) food, b) social and c) flower cues in the three different experimental sessions (Fasting, Baseline and Isolation), with black dots indicating the average NCS response for each participant in response to the given cue (N = 30). Experimental conditions hypothesized to increase NCS responses in a modality-specific way are highlighted in bold (fasting session for food cues and social isolation session for social cues). NCS responses to food cues were significantly higher in the fasting versus the baseline session (CI: 0.12, 1.28, t (29) = 2.46, *p* = 0.020, *d* = 0.45). NCS responses to social cues were significantly higher after social isolation versus baseline session (CI: 0.17, 1.58, t (29) = 2.53, *p* = 0.017, *d* = 0.46). There was no significant effect of session on NCS responses to flower cues.